\RequirePackage{fix-cm}
\documentclass[smallextended]{svjour3} 
\smartqed 
\usepackage{eso-pic}
\usepackage{graphicx}
\usepackage{booktabs}
\usepackage{marginnote}

\usepackage{bm}

\usepackage[utf8]{inputenc}
\usepackage[T1]{fontenc}
\usepackage{mathptmx}

\newcommand{\sss}{\scriptscriptstyle}
\newcommand{\sst}{\scriptstyle}

\newcommand{\stext}[1]{\sss \rm{#1} \sst}
\newcommand{\rzcm}{cm$^{-1}$}
\newcommand{\veps}{\varepsilon}
\usepackage{eso-pic}
\makeatletter

\begin{document}

\title{Terahertz to mid-infrared dielectric properties of polymethacrylates for stereolithographic single layer assembly}

\author{Serang Park$^{1}$\and
        Yanzeng Li$^{1}$\and
		Daniel B.~Fullager$^{2}$ \and
        Stefan Sch\"{o}che$^{3}$\and
		Craig M.~Herzinger$^{3}$\and
       	Glenn D.~Boreman$^{1}$ \and
        Tino Hofmann$^{1,4}$ }

\institute{Serang Park \at
\email{spark71@uncc.edu}\\
Phone: +1 704 687 8622 \\
              1) Department of Physics and Optical Science, University of North Carolina at Charlotte, 9201 University City Blvd, Charlotte, NC, 28223\at
              2) Lasertel, 7775 N Casa Grande Highway, Tucson, AZ, 85743\at
              3) J. A. Woollam Co. Inc., 645 M Street, Suite 102, Lincoln, NE 68508, USA\at
              4) THz Materials Analysis Center, Department of Physics, Chemistry, and Biology (IFM), Link{\"o}ping University, SE 581 83 Link{\"o}ping, Sweden\at
}

\date{Received: 07/18/2019}

\titlerunning{Terahertz to mid-infrared dielectric properties of polymethacrylates}
\journalname{J. Infrared, Milimeter, Terahertz Waves}

\maketitle

\begin{abstract}
\noindent The fabrication of terahertz (THz) optics with arbitrary shapes via poly-methacrylate-based stereolithography is very attractive as it may offer a rapid, low-cost avenue towards optimized THz imaging applications. In order to design such THz optical components appropriately, accurate knowledge of the complex dielectric function of the materials used for stereolithographic fabrication is crucial. In this paper we report on the complex dielectric functions of several polymethacrylates frequently used for stereolithographic fabrication. Spectroscopic ellipsometry data sets from the THz to  mid-infrared spectral range were obtained from isotropically cross-linked polymethacrylate samples. The data sets were analyzed using stratified layer optical model calculations with parameterized model dielectric functions. While the infrared spectral range is dominated by a number of strong absorption features with Gaussian profiles, these materials are found to exhibit only weak absorption in the THz frequency range. In conclusion, we find that thin transmissive THz optics can be efficiently fabricated using polymethacrylate-based stereolithographic fabrication.


\end{abstract}

\section{Introduction}
\label{intro}

THz optical components produced by additive manufacturing techniques have received considerable interest in recent years as cost effective rapid-prototyping solutions for THz optical components such as lenses \cite{squires20153d}, filters and waveguides \cite{Weidenbach:16,Kaur2015}. A large body of literature has reported on the performance and optical properties of THz optical components fabricated using 3D printing techniques \cite{busch2014optical,busch2016thz,furlan20163d,Park2013IEEE}.
Several publications are dedicated to the THz optical properties of the materials suitable for 3D printing \cite{busch2014optical,busch2016thz}. However, the major focus to date is on materials and optical components fabricated using fused deposition-based techniques \cite{squires20153d,Weidenbach:16,busch2014optical}. The advantages of fused deposition-based techniques are primarily the low instrument and fabrication costs as well as the large variety of compatible materials \cite{Yan1996}. The resolution and surface finish of fused deposition-based techniques are limited by the nozzle diameter through which the materials are applied. Current state-of-the-art fused deposition-based printers have nozzle diameters that range on the order of several hundred~$\mu$m \cite{Okwuosa2017,Zhang2017IEEE}. Stereolithography, in contrast, has been demonstrated to achieve resolution on the order of 10~$\mu$m and substantially better surface finish compared to other fabrication techniques \cite{NGO2018172,Shallan2014chem}. Therefore, stereolithography-based techniques are ideal methods of fabrication for THz optical components \cite{OtterPotI105_2017,Fullager2019}.

Despite the rapidly increasing interest in 3D printed THz and infrared optical components, accurate infrared and THz dielectric function data on polymethacrylates available for stereolithography-based fabrication have not been reported yet. This hinders the development of optical components composed of such polymethacrylates and impedes the progress in simulation-based design of metamaterials with novel THz optical properties.

In this paper we report on the first ellipsometric measurements and the complex dielectric functions of stereolithography-compatible polymethacrylates in the mid-infrared and THz spectral range. Three different commercially available polymethacrylates (Formlabs Inc.) were investigated. We find that the polymethacrylates exhibit very similar THz optical properties but display characteristic differences in the absorption bands observed in the infrared spectral range. In the THz spectral range all investigated materials show sufficient transparency to allow the fabrication of thin transmissive optical components, in particular for the lower THz frequency range. A parametrized model dielectric function composed of harmonic oscillators with Gaussian broadening is derived and discussed. 

\vspace{-0.2cm}
\section{Experiment}
\label{sec:exp}
\vspace{-0.1cm}
\subsection{Sample Preparation}
\label{sec:prep}
\vspace{-0.2cm}
The samples studied here were prepared using UV-induced polymerization of the methacrylate-based resins in a mold as described below. This fabrication process ensured that the surface roughness is sufficiently small for infrared ellipsometry. For each sample, approximately 2~ml of resin was applied in between two microscope slides placed parallel to each other on a glass plate. Subsequently, a second glass plate was set on top of the spacers to shape the resin into a thin slab. The assembly was then placed in a UV oven (UVO cleaner model no.~42, Jelight Company Inc.) and was cured for 15 minutes until the resin was fully polymerized. As a result, the polymerized resin samples have parallel interfaces with low surface roughness and are suitable for accurate ellipsometric measurements in the infrared and THz spectral range. This fabrication approach was applied for three different methacrylate-based resins, which are commercially available (Formlabs Inc.) described by the vendor as ``castable'' (sample~1),``tough'' (sample~2), and ``black'' (sample~3). All investigated samples have a nominal thickness of 1~mm.

\subsection{Data Acquisition and Analysis}
\label{sec:ana}
The polymethacrylate samples were investigated using a commercial infrared ellipsometer (Mark I IR-VASE$^{\textregistered}$, J.A. Woollam Company Inc.) and a commercial THz ellipsometer (THz-VASE, J.A. Woollam Company Inc.). The IR ellipsometer operates in a polarizer ${-}$ sample ${-}$ rotating compensator ${-}$ analyzer configuration, while the THz ellipsometer uses a rotating polarizer ${-}$ sample ${-}$ rotating compensator ${-}$ analyzer configuration as detailed in Ref.~\cite{Fujiwara2007}. The infrared ellipsometer is equipped with a Fourier transform infrared (FTIR) spectrometer and employs a deuterated-triglycine sulfate (DTGS) detector. The THz ellipsometer is equipped with backward-wave oscillator source operating in the range from 100 to 180~GHz. Schottky diode frequency multipliers are used to extend the spectral range from 0.65 to 0.95~THz. A Golay cell is employed as a detector in the THz-VASE. Ellipsometric $\Psi$- and $\Delta$-spectra were obtained in the infrared spectral range from 300 to 4000~cm$^{-1}$ (9 to 120~THz) with a resolution of 4~cm$^{-1}$ (0.1~THz) at three angles of incidence: $\Phi _{a}$ = 65$^{\circ}$, 70$^{\circ}$, and 75$^{\circ}$. The THz ellipsometric data were obtained over range from 22 to 32~cm$^{-1}$ (0.65 to 0.95~THz) with a resolution of 0.2~cm$^{-1}$ (5~GHz) at the same angles of incidence as for the infrared data.

The optical modeling and data analysis were performed using a commercial ellipsometry data analysis software package (WVASE32$^\textsuperscript{TM}$, J.A. Woollam Company). The complete ellipsometric data set obtained for each sample was analyzed using a three layer optical model composed of air ${-}$ polymethacrylate ${-}$ air. A model dielectric function was used to describe the infrared and THz optical response of the polymethacrylates. The model dielectric function incorporates a sum of Gaussian oscillators:

\begin{equation}
\label{eq:eps_all}
\varepsilon(\omega) = \varepsilon_{1}(\omega)+i \varepsilon_{2}(\omega) = \veps_{\infty} + \sum_{i} \varepsilon_{Gau}({A}, \Gamma, \omega, \omega_o),
\end{equation}

\noindent where the function $\varepsilon_{Gau}({A}, \Gamma, \omega, \omega_o)$ indicates an oscillator with Gaussian broadening. The oscillator amplitude, broadening, and resonance frequency are designated by $A, \Gamma$, $\boldmath \omega_o$, respectively. The oscillators are given analytically by their Gaussian form for the imaginary part $\varepsilon_{2}^{\stext{Gau}} (\omega)$ of the complex dielectric function $\varepsilon(\omega)$:

\begin{equation}
\label{eq:Gaussian}
\varepsilon_{2}^{\stext{Gau}} (\omega) = A\exp{\left(-\left(\frac{\omega - \omega_o}{f\cdot\Gamma}\right)^{2}\right)} +  A\exp{\left(-\left(\frac{\omega + \omega_o}{f\cdot\Gamma}\right)^{2}\right)},
\end{equation}
 
\noindent where $1/f = 2\sqrt{\rm{ln}(2)}$. The corresponding values for $\varepsilon_{1}^{Gau}(\omega)$ are determined by Kramers-Kronig integration of Eq.~(\ref{eq:Gaussian}) during the Levenberg-Marquardt-based lineshape analysis of the experimental $\Psi$- and $\Delta$-spectra. 

During the lineshape analysis, relevant model parameters are varied until the best match between calculated model and experimental ellipsometry data is achieved. The best-model calculated spectra shown in Figs.~\ref{fig:psi} and \ref{fig:del} require 14 Gaussian oscillators with frequencies ranging from 40 to 3500~\rzcm\ for sample~1. The analysis for samples~2 and 3 required a model dielectric function composed of 15 distinct Gaussian oscillators in this energy range in order to describe the experimentally observed lineshapes.

\section{Results and Discussion}
\label{sec:res}

\begin{figure}[b!]
	\centering
	\includegraphics[width=0.7\linewidth, trim=0 400 0 20,clip]{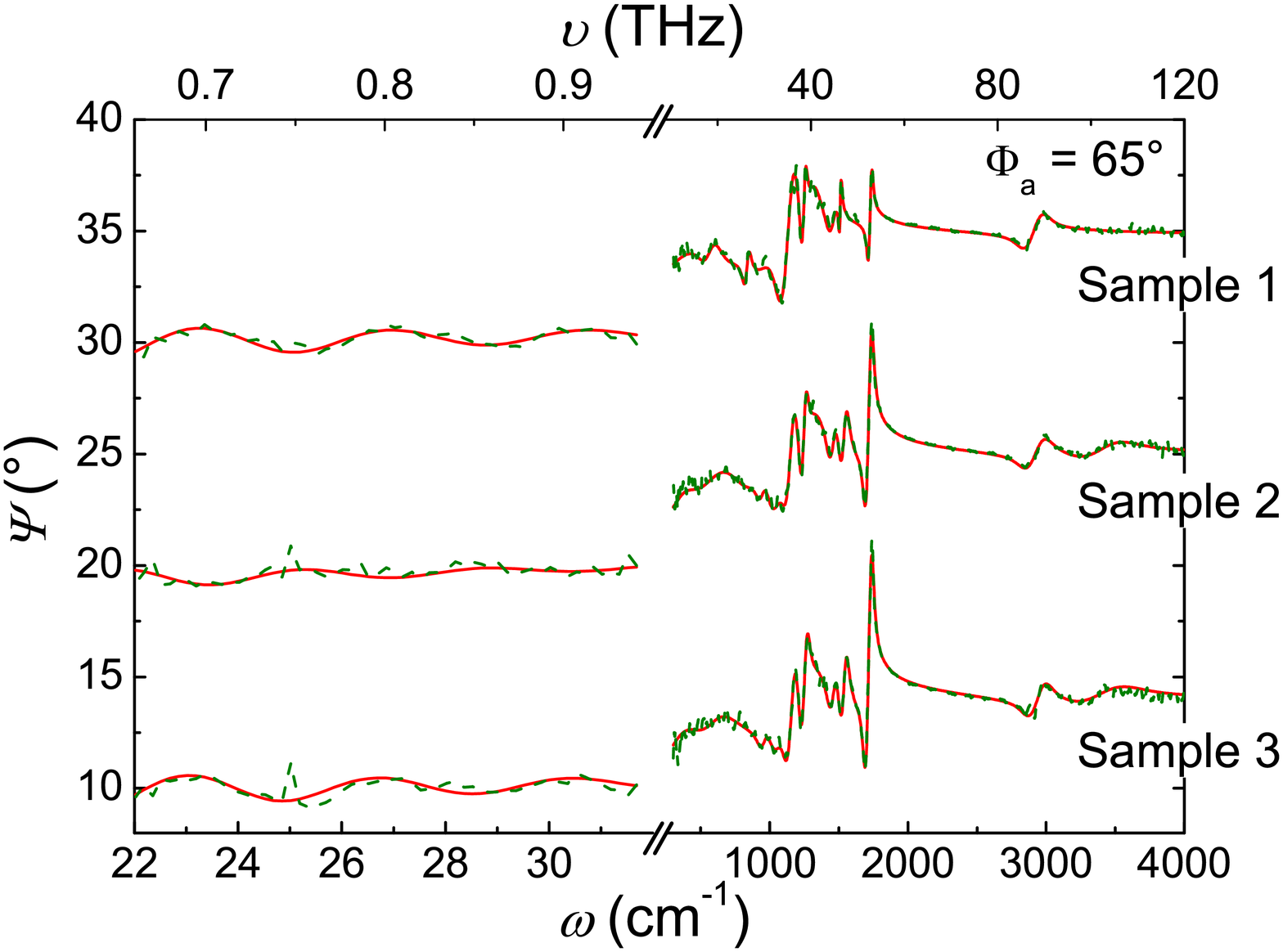}
	\caption{Best-model calculated (red solid lines) and experimental (dashed green lines) $\Psi$-spectra obtained at $\Phi _{a}$ = 65$^{\circ}$ for sample 1, 2, and 3. The infrared range is dominated by a number of distinct absorption bands while the THz range shows Fabry-P\'erot oscillations as a result of the plane parallel interfaces of the samples. The $\Psi$-spectra of sample~2 and 3 are shifted with respect to the $\Psi$-spectrum of sample 1 by a constant offset of 12$^{\circ}$ and 22$^{\circ}$, respectively.}
	\label{fig:psi}      
\end{figure}

\begin{figure}[ht!]
	\centering
	\includegraphics[width=0.7\linewidth, trim=0 400 0 20,clip]{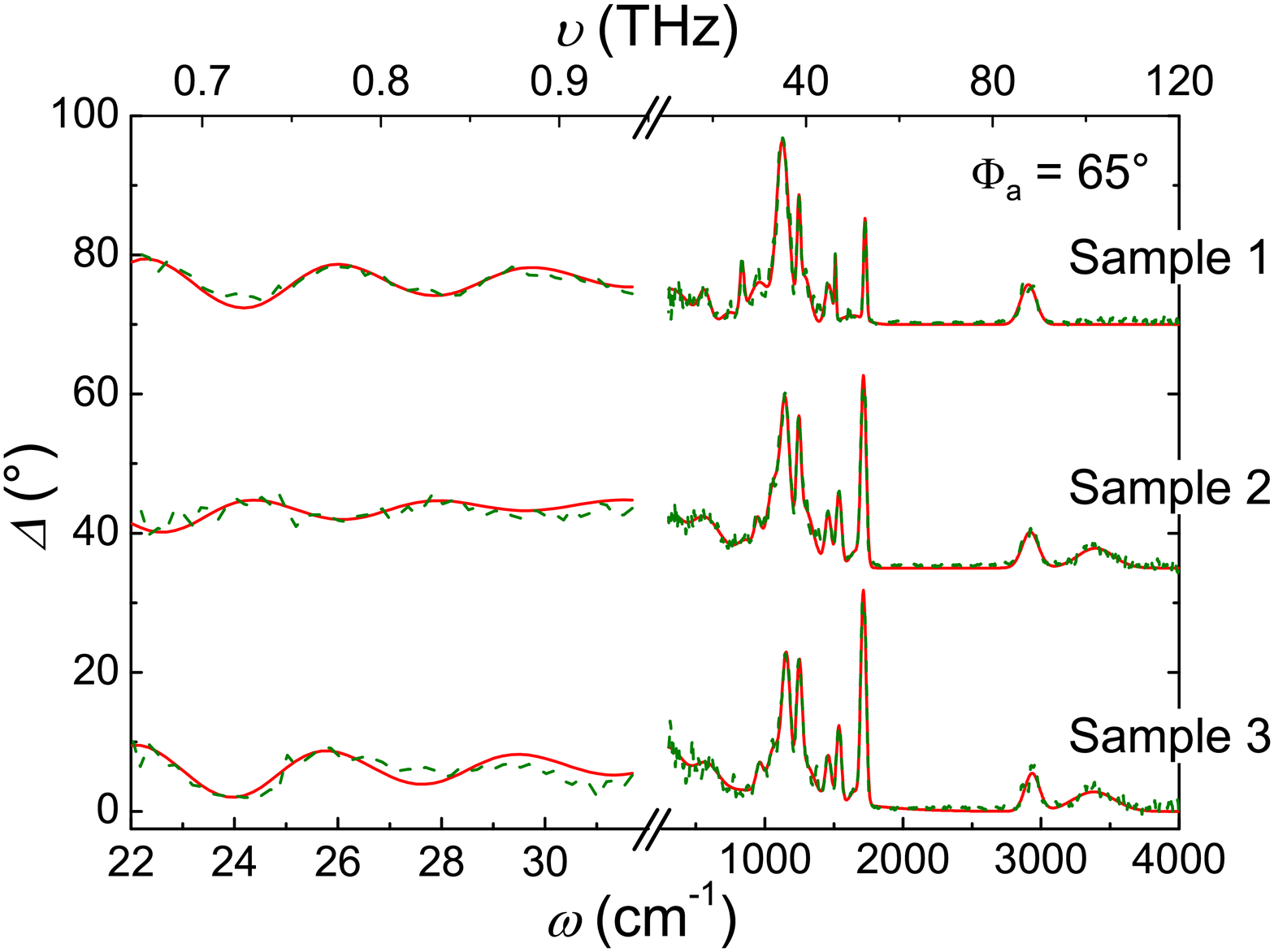}
	\caption{Best-model calculated (red solid lines) and experimental (dashed green lines) $\Delta$-spectra obtained at $\Phi _{a}$ = 65$^{\circ}$ for sample 1, 2, and 3. Similar to the $\Psi$-spectra shown in Fig.~\ref{fig:psi}, the infrared range is dominated by several distinct absorption bands while the THz range exhibits Fabry-P\'erot oscillations as a result of the plane parallel interfaces of the samples. The $\Delta$-spectra of sample~2 and 3 are shifted with respect to the $\Delta$-spectrum of sample~1 by a constant offset of 35$^{\circ}$ and 70$^{\circ}$, respectively.}
	\label{fig:del}       
\end{figure}

\noindent Figure \ref{fig:psi} illustrates the experimental (dashed green lines) and the  best-model calculated (solid red lines) $\Psi$-spectra of all three samples at the angle of incidence $\Phi _{a}$ = 65$^{\circ}$ for the spectral range from 22 to 4000~\rzcm. The corresponding $\Delta$-spectra are shown in Fig.~\ref{fig:del}. Note that experimental $\Psi$- and $\Delta$-spectra were obtained for $\Phi _{a} = 65^{\circ}, 70^{\circ},$ and $75^{\circ}$ and were analyzed simultaneously. Only the data for $\Phi _{a}$ = 65$^{\circ}$ is shown here for clarity. The experimental and best-model calculated data are in very good agreement for all investigated samples. All three samples show a very similar THz and infrared response where Fabry-P\'erot oscillations are observed in the range from 22 to 32~cm$^{-1}$ (0.65 to 0.95~THz) and distinct absorption bands are found in the range from 300 to 4000~cm$^{-1}$ (9 to 120~THz). 

A subtle damping in the Fabry-P\'erot oscillations in Fig.~\ref{fig:psi} and \ref{fig:del} can be noticed. This is accounted for in the model with a broad, shallow oscillator at $\omega$ = 40~cm$^{-1}$. While the Fabry-P\'erot oscillations of samples~1 and 3 have very similar amplitudes, the amplitudes of the Fabry-P\'erot oscillations for samples~2 are distinctly smaller, indicating a lower transparency in this spectral range for sample~2. 

Although very similar at a first glance, the investigated polymethacrylates show subtle differences in the mid-infrared spectral range from $\omega=500$ to 2000~cm$^{-1}$. In this spectral range, sample~1 exhibits a peak at approximately 800~cm$^{-1}$, which is not observed in sample~2 and 3. At higher wavenumbers, at approximately 3400~cm$^{-1}$, samples~2 and 3 show a distinct absorption peak that is not present in sample~1. Thus, sample~1 can be easily distinguished from the other investigated polymethacrylate samples by its unique infrared fingerprint. The polymethacrylates in sample~2 and 3, however, show a very similar response in the infrared spectral range. Thus, a differentiation between the two materials in this spectral range requires quantification of oscillator amplitude and broadening parameters.

\begin{figure}[]
	\centering
	\includegraphics[width=0.7\linewidth, trim=0 450 0 0,clip]{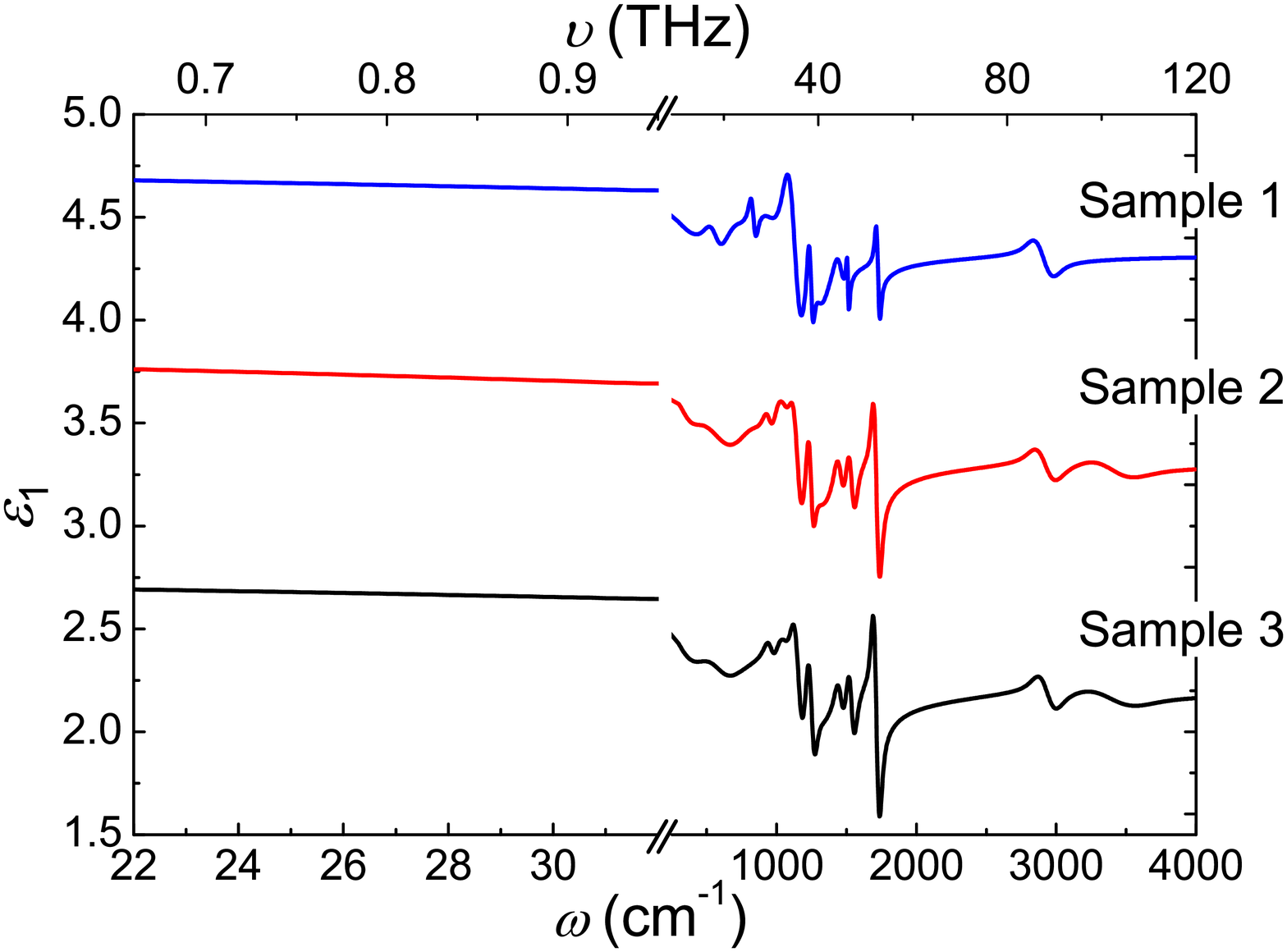}
	\caption{Best-model calculated real part of the complex dielectric function $\varepsilon(\omega)$ for sample 1, 2, and 3 are shown in solid lines. The major contributions to the dispersive behavior of all three samples occur in spectral range from 300 to 4000 cm$^{-1}$. The best-model parameters of the model dielectric function are given in Tabs.~\ref{tab:param1} and \ref{tab:param2}. Note that the $\varepsilon_{1}(\omega)$-spectra of sample~2 and 1 are shifted with respect to the $\varepsilon_{1}(\omega)$-spectrum of sample~3 by a constant offset of 1 and 2, respectively.}
	\label{fig:e1}       
\end{figure}

\begin{figure}[]
	\centering
	\includegraphics[width=0.7\linewidth, trim=0 450 0 0,clip]{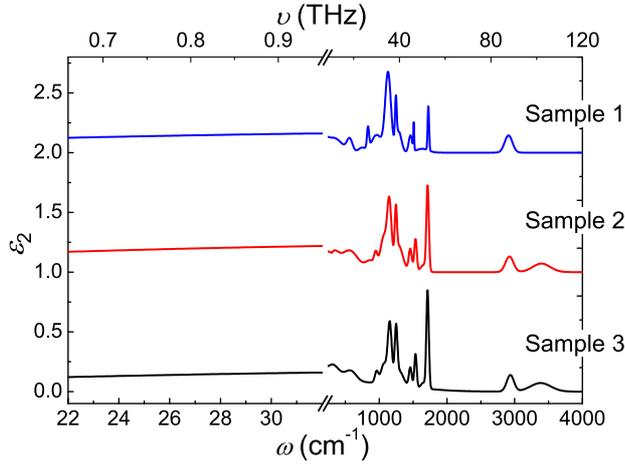}
	\caption{Best-model calculated imaginary part of the complex dielectric function $\varepsilon(\omega)$ for sample 1, 2, and 3 are shown in solid lines. The major contributions to the absorptive behavior of all three samples occur in infrared spectral range from 300 to 4000 cm$^{-1}$, while only a broad and shallow absorption was observed throughout the THz range. The best-model parameters are given in Tabs.~\ref{tab:param1} and \ref{tab:param2}. Note that the $\varepsilon_{2} (\omega)$ spectra of sample~2 and 1 are shifted with respect to the $\varepsilon_{2} (\omega)$ spectrum of sample~3 by a constant offset of 1 and 2, respectively.}
	\label{fig:e2}       
\end{figure}

Figs.~\ref{fig:e1} and \ref{fig:e2} show the real and imaginary parts of the model dielectric functions $\varepsilon(\omega)$ of all three samples for comparison. As discussed, all samples show similar responses in the infrared and THz ellipsometric data shown in Figs.~\ref{fig:psi} and \ref{fig:del} and correspondingly, the model dielectric functions of these samples are very similar. Strong absorption bands can be identified in the range from $\omega = 500$ to 2000~cm$^{-1}$. This allows a fingerprint identification of the materials in this spectral range.

In the THz spectral range from 0.65 to 0.95~THz (22 to 32~\rzcm) the permittivity for the samples shows little dispersion and absorption, which results in the Fabry-P\'erot oscillations in these layers with plane parallel interfaces as seen in Figs.~\ref{fig:psi} and \ref{fig:del}. Sample~1 shows the smallest $\varepsilon_{2}(\omega)$ while sample~2 shows the largest $\varepsilon_{2}(\omega)$, which makes the material in sample~1 (``castable'', Formlabs Inc.) more suitable for transmissive optics. In comparison with the materials commonly used for fused deposition-based fabrication techniques, the absorption coefficients of the investigated materials are comparable. All samples exhibit absorption coefficient values larger than 10~cm$^{-1}$ over the range from 0.65 to 0.95~THz, which is similar to that of nylon and polylactic acid 90. Polystyrene and butadiene over the same spectral range exhibit values below 5~cm$^{-1}$ \cite{busch2014optical}. Here we account for the absorption in this spectral range by using a broad and shallow Gaussian oscillator, which is located in the spectral gap between operational ranges of the THz and infrared ellipsometers. Its resonance frequency was approximated at $\omega$ = 40~cm$^{-1}$, which was not further varied during the model analysis. 

\noindent Tables \ref{tab:param1} and \ref{tab:param2} summarize the best-model oscillator parameters for the dielectric response of the polymethacrylates in sample~1, 2, and 3 for comparison. The oscillator frequency $\omega_o$ and broadening $\Gamma$ are given in units of cm$^{-1}$, while amplitude $A$ is dimensionless. Error bars in parentheses represent the 90\% confidence limits of the respective oscillator parameters. 

\begin{table}[ht]
	\footnotesize
	\centering
	\caption{Comparison of the best-model oscillator energies of the absorption bands identified for samples 1-3 in the range from 22 to 4000~\rzcm (0.65 to 120~THz). The parentheses indicate a 90\% confidence interval for the corresponding digits.}
	\label{tab:param1}
	\begin{tabular}{cccc}
		
		\hline\hline
		& \multicolumn{1}{c}{Sample~1}		 & \multicolumn{1}{c}{Sample~2}		    & \multicolumn{1}{c}{Sample~3}		 \\
		\cmidrule(r){2-2}						\cmidrule(r){3-3}						\cmidrule(r){4-4}
		Number			& 	$\omega_o$  (\rzcm)	  		 & 		$\omega_o$  (\rzcm)	 			& 	$\omega_o$  (\rzcm)	 		 	\\
		\hline
		1 				&40* 							 &40*				 					&40*								\\
		2 				&318	(28)  					 &322	(29)							&292	(18)						\\
		3 				&564	(6) 	 				 &557	(29) 		 					&578	(23)						\\
		4 				&744	(13)  					 &874	(23) 		 					&739	(314)					   	\\
		5 				&834	(1) 	 				 &949	(3)								&961	(4) 					 	\\
		6 				&962	(4) 	 				 &1070	(9) 		 					&1070	(7)							\\
		7 				&1130	(1)  					 &1149	(1) 		 					&1156	(1)						   	\\
		8 				&1245.7	(4) 					 &1245	(1)								&1249	(1)						 	\\
		9 				&1286	(4)  					 &1281	(7) 		 					&1297	(21)						\\
		10 				&1461	(1)  					 &1460	(1) 		 					&1459	(1)						   	\\
		11 				&1510.3	(4) 					 &1536	(1)								&1535.8	(4)							\\
		12 				&1635	(13) 					 &1655	(11)		 					&1648	(7)							\\
		13 				&1725.5	(3) 					 &1725.5 (3)			 				&1713.6	(2)						   	\\
		14 				&2909	(2) 					 &2922	(2)								&2935	(2)						 	\\
		15 				&- 								 &3395	(7) 		 					&3382	(7)							\\
		\hline\hline
	\end{tabular}\\
	{* Value} not further varied during the fit process.
\end{table} 

\begin{table}[h!]
	\footnotesize
	\centering
	\caption{Comparison of the best-model parameters for the amplitude and broadening of the Gaussian oscillators identified for samples~1-3 with energies as shown in Tab.~\ref{tab:param1} in the range from 22 to 4000~\rzcm (0.65 to 120~THz). The parentheses indicate a 90\% confidence interval for the corresponding digits.}
	\label{tab:param2}
	\begin{tabular}{ccccccc}
		
		\hline\hline
						& \multicolumn{2}{c}{Sample~1}		 & \multicolumn{2}{c}{Sample~2}		    & \multicolumn{2}{c}{Sample~3}		 \\
\cmidrule(r){2-3}\cmidrule(r){4-5}\cmidrule(r){6-7}
		Number			& 	$A$			 & $\Gamma$  (\rzcm)&$A$	  		 & 	$\Gamma$ (\rzcm)&$A$	  		&$\Gamma$  (\rzcm)\\
		\hline
		1 				&0.164	(14)	 &59	(61) 		&0.234	(19)     &59	(49)		&0.164	(14) 	&56		(40)\\
		2 				&0.125	(20)	 &306	(86) 	 	&0.14	(5) 	 &172   (84) 		&0.186	(29) 	&279	(65)\\
		3 				&0.103	(19)	 &95	(17) 		&0.182	(9) 	 &310	(87)		&0.098	(19) 	&199	(37)	\\
		4 				&0.039	(6)	 	 &120	(39) 		&0.088	(15)     &174	(73)	 	&0.081	(18)	&1474	(305)	\\
		5 				&0.177	(12)	 &36	(3) 		&0.107	(32) 	 &57	(11)	  	&0.094	(10)	&59		(8)	\\
		6 				&0.149	(4) 	 &179	(16) 		&0.297	(17)	 &121	(20)		&0.162	(49)	&102	(20)	\\
		7 				&0.665	(7) 	 &99.5  (18) 	 	&0.508	(65)     &70	(5)	 		&0.495	(33)	&68		(4) 	\\
		8 				&0.354	(1)		 &29.3  (13) 		&0.366	(19)	 &37	(2)	  		&0.423	(40) 	&46		(3) \\
		9 				&0.176	(6) 	 &98	(7) 	 	&0.225	(9) 	 &158	(12)		&0.117	(14) 	&140	(34)	\\
		10 				&0.140	(6)		 &61	(4) 		&0.191	(6)      &55	(3)	 		&0.160	(7) 	&50		(3) 	\\
		11 				&0.231	(1)		 &13	(1) 		&0.275	(6) 	 &45	(1)	  		&0.279	(7) 	&41		(1) \\
		12				&0.032	(3) 	 &210	(28) 	 	&0.060	(5) 	 &103	(20)		&0.044	(5) 	&72		(16)	\\
		13 				&0.372	(8)		 &24.7  (7) 	 	&0.704	(16)     &42	(1)	 		&0.821	(8) 	&42.5	(5) 	\\
		14 				&0.144	(4)  	 &133	(4) 		&0.1304	(4)		 &139	(5)	  		&0.138	(4) 	&124	(4) \\
		15 				&-				 &-					&0.072	(4)		 &305	(17)		&0.072	(3) 	&353	(18)\\
		\hline\hline
	\end{tabular}\\
\end{table}

\section{Summary and Conclusion}
\label{summary}
In this work we report accurate complex dielectric function values of three different polymethacrylates that are available for stereolithography-based manufacturing with commercial 3D printers in the infrared and THz spectral range. A model dielectric function composed of multiple oscillators with Gaussian broadening was found to appropriately render the THz and infrared ellipsometric responses. While the investigated materials are transparent in the THz spectral range, the infrared spectral range is dominated by distinct absorption bands. In contrast to materials commonly used for fused deposition-based fabrication techniques, such as acrylonitrile butadiene styrene or polystyrene, polymethacrylates analyzed here exhibit relatively high absorption across THz range \cite{busch2014optical}. However they are still sufficiently transparent for the fabrication of thin transparent THz optics, such as Fresnel lenses. We anticipate that the parametrized dielectric functions reported here will help to improve first-principle calculations of the infrared and THz optical responses of 2D and 3D structures composed of these materials.

\begin{acknowledgements}
	SP, YL, and TH would like to acknowledge the valuable discussions with Susanne Lee and Erin Sharma within the NSF IUCRC for Metamaterials. The authors are grateful for support from the National Science Foundation (1624572) within the I/UCRC Center for Metamaterials, the Swedish Agency for Innovation Systems (2014-04712), and Department of Physics and Optical Science of the University of North Carolina at Charlotte.
\end{acknowledgements}

\end{document}